\documentclass[graybox]{llncs}

\usepackage{mathptmx}       
\usepackage{helvet}         
\usepackage{courier}        
\usepackage{type1cm}        
\usepackage{makeidx}         
\usepackage{graphicx}        
\usepackage{multicol}        
\usepackage[bottom]{footmisc}

\makeindex             

\usepackage{textcomp}	

\begin{document}

\title{Fuzzy-Based Forest Fire Prevention and Detection by Wireless Sensor Networks }
\author{Josu\'e Toledo-Castro, Iv\'an Santos-Gonz\'alez, Pino Caballero-Gil, Candelaria Hern\'andez-Goya, Nayra Rodr\'iguez-P\'erez, Ricardo Aguasca-Colomo}

\institute{ Josu\'e Toledo-Castro, University of La Laguna, \email{alu0100763492@ull.edu.es}
\and Iv\'an Santos-Gonz\'alez, University of La Laguna, \email{jsantosg@ull.edu.es}
\and Pino Caballero-Gil, University of La Laguna \email{pcaballe@ull.edu.es}
\and Candelaria Hern\'andez-Goya, University of La Laguna, \email{mchgoya@ull.edu.es}
\and Nayra Rodr\'iguez-P\'erez, University of La Laguna \email{mrodripe@ull.edu.es}
\and Ricardo Aguasca-Colomo, IUSIANI - ULPGC, \email{ricardo.aguasca@ulpgc.es}}

%
%
\maketitle

\abstract{
Forest fires may cause considerable damages both in ecosystems and lives. This proposal describes the application of Internet of Things and wireless sensor networks jointly with multi-hop routing through a real time and dynamic monitoring system for forest fire prevention. It is based on gathering and analyzing information related to meteorological conditions, concentrations of polluting gases and oxygen level around particular interesting forest areas. Unusual measurements of these environmental variables may help to prevent wildfire incidents and make their detection more efficient. A forest fire risk controller based on fuzzy logic has been implemented in order to activate environmental risk alerts through a Web service and a mobile application. For this purpose, security mechanisms have been proposed for ensuring integrity and confidentiality in the transmission of measured environmental information. Lamport's signature and a block cipher algorithm are used to achieve this objective.
}


\section{Introduction}

Huge losses may be caused as consequence of emergency situations and natural disasters such as forest fires implicating serious threats to the ecosystems and people's health. The lack of real time systems that allow managing forest resources and activating wildfire incident alerts may cause difficulties in forest fires prevention, detection and fighting. Some environmental events and variables considered as dynamic forest fire risk factors, such as meteorological variables, polluting gases or oxygen, can be used in order to estimate an environmental risk index of wildfire incidents. Their real time monitoring over different forest areas may favor efficiency on the response time of emergency bodies, helping to prevent forest fires, notifying their detection or tracking their progress.

To this end, the use of Internet of Things (IoT) devices and sensors can be relevant for developing Wireless Sensor Networks (WSN) \cite{kumar2014wireless} for environmental protection. Regarding their deployment in outdoor forest areas, some challenges such as communication among distributed nodes, possible areas out of network coverage or the implementation of security mechanisms should be considered.

Thus, this proposal is devoted to implementing a real time monitoring system for different environmental variables through a wireless IoT sensor network distributed over different forest areas aimed to prevention, detection and tracking of wildfire incidents. For this purpose, a forest fire risk controller based on fuzzy logic \cite{zadeh2015fuzzy} has been integrated. It is responsible for interpreting and analysing whether measurements of meteorological variables, polluting gases and oxygen registered by IoT sensors may evidence forest fire risks. In order to activate and manage environmental alerts and information, a Web service and a mobile application are included. Particular attention has been paid to the system security. Secure communications among the IoT devices and the Web service and mobile devices, based on  multi-hop routing \cite{rani2015multi}, has been developed. Lamport's signature scheme \cite{merkle1989certified}, a block cipher algorithm and a secure authentication scheme have been implemented as well.

This work is organized as follows. Section 2 deals with the description of some related works. Then, the method applied to evaluate the existence of forest fire risks is detailed in Section 3 and the proposed system is outlined in Section 4. Finally, Section 5 explains the implemented security mechanisms and Section 6 provides conclusions, research works in progress and future research lines.

\section{Related Works}
Currently, many proposals have been implemented with the aim of offering environmental monitoring solutions based on fuzzy logic. In this sense, the work \cite{rajkumar2017iot} proposes a smart system to analyse carbon dioxide emissions in the same way that the paper \cite{zhou2011identifying}, both combine fuzzy logic and decision-making trial to improve efficiency in emergency management. To this respect, the paper \cite{iliadis2005decision} suggests a long-term forest risk estimation for determining the existence of high or low risk in Mediterranean basin countries through applying an integrated fuzzy model together with machine learning.

Another relevant aspect regarding WSN, IoT and multi-hop routing is the approach to consider communication security as a main goal. In this sense, the implementation of key predistribution schemes should be considered. The paper \cite{chan2003random} provides three different mechanisms for key establishment through pre-distributing a random set of keys to each node. Others security issues and challenges need to be also considered, the work \cite{pathan2006security} identifies security threats.

The system proposed here presents the use of wireless sensor networks, fuzzy logic, multi-hop routing and security mechanisms for developing a secure environmental measurement interface, able to measure dynamic risk factors that make forest fire prevention, detection and tracking more efficient. It also provides a forest fire risks controller for evaluating recent wildfire incidents. It is composed by a Web service, a mobile application and a particular IoT device containing environmental sensors.
 
\section{Method}

A wireless sensor network based on IoT devices and environmental sensors is deployed through forest areas. The objective is to perform a continuous environmental monitoring of dynamic risk factors for forest fires such as meteorological variables (temperature, relative humidity, wind speed and rainfall), polluting gases (such as carbon dioxide and monoxide) and oxygen level. These variables are measured on every forest area and the information gathered is transmitted to a Web service in charge of analysing the possible existence of forest fire risks as well as any evidence of recent wildfire incidents.

\begin{figure}[htb!]
  \centering
    \includegraphics[scale=0.18]{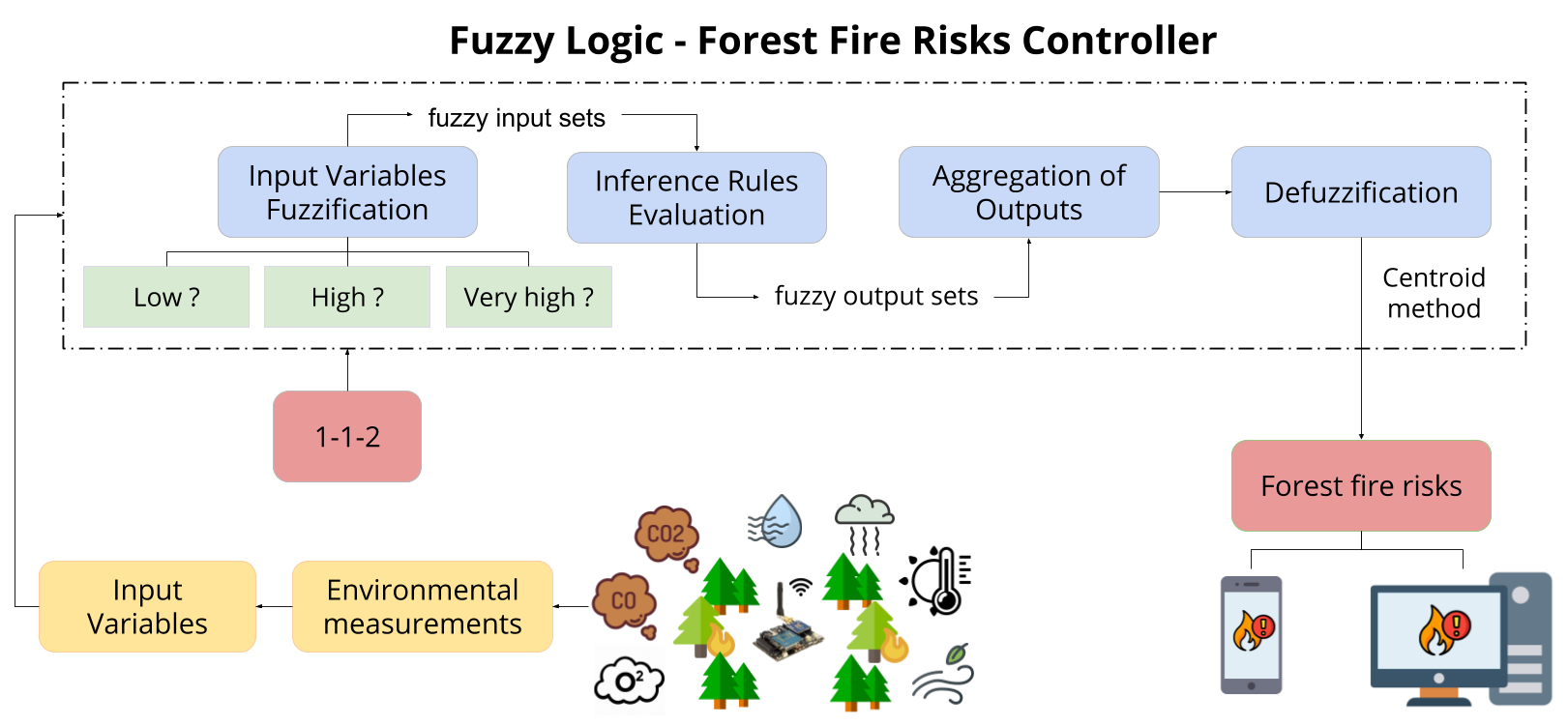}
  \caption{Fuzzy logic - Forest Fire Risks Controller}
  \label{fig:fuzzylogic}
\end{figure}

Estimating forest fire risk factors is not simple and cannot be executed with complete accuracy. For this purpose, fuzzy logic and Mamdani Inference \cite{Mamdani} have been used here in order to obtain a forest fire environmental risk controller. These methods allow to include the uncertainty of environmental data, the imprecision related to the variation in the parameters and the behavior of every monitored variable (see Figure \ref{fig:fuzzylogic}).

\subsection{Linguistic variables}

The environmental variables aforementioned are considered as the input variables of the proposed fuzzy logic system.

Meteorological variables set includes temperature, relative humidity, wind speed and rainfall as linguistic variables. Their fuzzy sets have been mainly proposed taking into account the ``Rule of 30" \cite{lecina2014extreme} as forest fire prevention model. It considers temperature measurements above 30 \textdegree C, humidity percentages below 30 \%, wind speed values above 30 km/h and rainfall measurements below 30 mm as relevant environmental conditions that may favor the occurrence of wildfire incidents. As consequence of increasing temperature and wind speed measurements and decreasing humidity percentages or rainfall measure above or below these thresholds, the severity of forest fire risks progressively increases from low to extreme risks. Every meteorological variable has as discourse of universe with its corresponding unit of measurement (Celsius degrees, percentages, kilometers per hour, etc.).

\begin{figure}[ht!]
  \centering
    \includegraphics[scale=0.27]{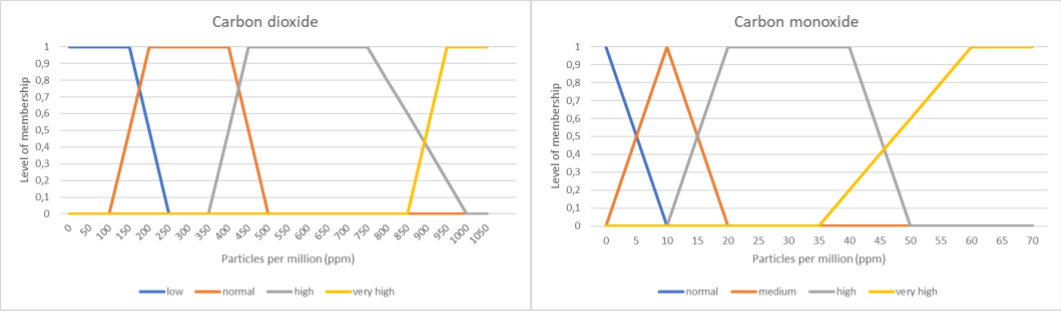}
  \caption{Example of fuzzy sets proposed for polluting gases}
  \label{fig:fuzzy}
\end{figure}

Regarding polluting gases, carbon dioxide and monoxide are also considered as input linguistic variables. These gases are produced in high levels during forest fires, so monitoring atypical increments in the carbon dioxide level (currently measured between 200 - 400 ppm) and the carbon monoxide (concentrations around 1 ppm can be commonly measured at outdoor areas) have to be considered as evidences of  environmental measurements at indoor places or consequence of performing industrial activities instead of outdoor forest areas \cite{Webpage:2017:maximaCO22017}. Taking into account these thresholds, fuzzy sets have been proposed (see Figure \ref{fig:fuzzy}). In both cases, particles per million (ppm) are considered as discourse of universe.

Finally, the oxygen level is also considered as input linguistic variable and shows an opposite behavior when comparing with polluting gases behavior. It means that instead of performing unusual increasing, oxygen level decreases progressively as consequence of being progressively consumed by fire. Hence, this variable may also be cataloged as a useful indicator. Despite the changes of oxygen level occurred over last millions of years, the current level may be defined as around 21\%. Fire needs at least 16 \% of oxygen level to occur. In contrast to polluting gases, percentages are used as discourse of universe. In this sense, all the fuzzy sets for every input variable have been proposed according to the knowledge of the experts that have been consulted.

In addition to this, the averages of environmental variables (temperature, humidity, carbon dioxide, etc.)  are also used as input linguistic variables.
Average calculation depends on prior detected forest fire risks for the current day in every monitored forest area. Taking into account every input linguistic variable, the corresponding average is calculated through all the previously registered measurements (when forest fire risks were not detected until that moment) and, on the other hand, the 15, 10 and 5 latest registered measurements (when low, high or extreme forest fire risk was previously detected in the same area). These averages are compared with the corresponding last registered measurement while evaluating inference rules. Likewise, reducing the measurements number from 15 to 5  for average calculation is also considered when external risk declarations by the emergency services are performed.

\subsection{Inference rules evaluation}

The inference rules proposed are based on evaluating the severity of forest fire risks (nonexistent, low, high or extreme) according to unusual variations between the last registered measurement and the average of every monitored environmental variable. In this way, detecting variations according to the proposed fuzzy sets of a particular variable between its last registered measurement and the average, may help to detect environmental conditions that have recently worsened or evidences of wildfire incidents occurrence. Every proposed inference rule evaluates these fuzzified input values with the aim of determining the fuzzified output result, that is represented on the proposed output variable membership function taking into account a percentage between 0\% and 100\% of detected forest fire risks.

For this purpose, Fuzzy Associative Memory (FAM) \cite{kosko1991fuzzy} is used as representation tool of the inference rules for evaluating the severity of forest fire risks with respect to the last registered measurement of every monitored environmental variable and its corresponding calculated average. Table \ref{tabla:FuzzyAssociativeMemory} shows the sample FAM proposed for the linguistic variable of carbon dioxide:

\begin{table}[ht]
\begin{center}
\begin{tabular}{|l|l|l|l|l|}
\hline
Last CO2 measurement / CO2 average & normal & medium & high &  very high \\ \hline
Normal & NFR & NFR & NFR & NFR \\ \hline
Medium & LFR & LFR & HFR & EFR \\ \hline
High  & HFR & HFR & HFR & EFR \\ \hline
Very High  & EFR & EFR & EFR & EFR \\ \hline
\end{tabular}
\caption{Inference rules evaluation for carbon dioxide variable}
\label{tabla:FuzzyAssociativeMemory}
\end{center}
\end{table}

In this sense, the comparison between carbon dioxide average and its last registered measurement may return nonexistent forest fire risks (NFR), low risks (LFR), high risks (HFR) or extreme forest fire risks (EFR) in the forest area where measurements were registered. This average is calculated depending on the forest fire risks previously declared. Likewise, this format based on FAM is used to represent the evaluation process of the inference rules for the rest of the linguistic variables (temperature, humidity, wind speed, rainfall, carbon monoxide and oxygen).

\subsection{Output variable}

Finally, the severity of detected fire risks in a particular forest area  is defined as the output linguistic variable. To this respect, the percentage has been considered as its discourse of universe.

Taking into account the evaluation between averages and the last measurements of every input variable, the inference rules evaluation returns several outputs always ranged from 0 \% to 100 \% and fuzzified into the output membership function. Before obtaining a discrete percentage of forest fire risks, the results of rules must be added together to generate the  output set (Aggregation of outputs step) and defuzzified through Centroid method \cite{runkler1997selection}.  According to the expert knowledge consulted, the triggers of the rules on the variables are previously set through appropriate overlaps of the fuzzy sets of input variables.

\begin{figure}[ht!]
  \centering
    \includegraphics[scale=0.11]{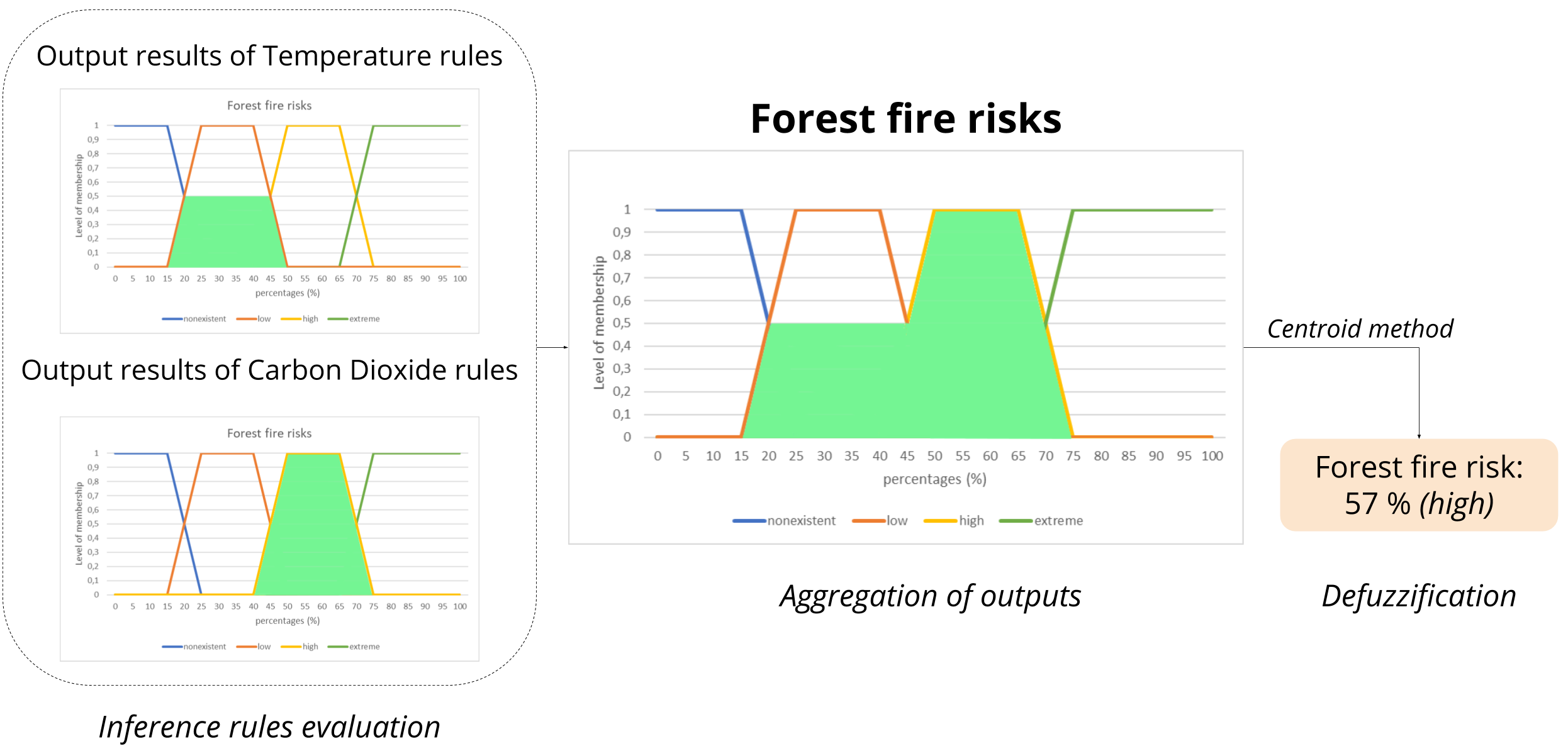}
  \caption{Aggregation of outputs and Defuzzification steps}
  \label{fig:dezzufication}
\end{figure}

For example, the Figure \ref{fig:dezzufication} shows the aggregation of the output results of inference rules evaluation on temperature and carbon dioxide input values into the same output set (previously fuzzified through their corresponding membership functions) in order to be defuzzified and express a discrete percentage for forest fire risks. The results of inference rules evaluated on the rest of input linguistic variables are also aggregated into the same final output set.

If forest fire environmental risks are detected by the proposed fuzzy system, the Web service is in charge of updating the environmental information  and activating new alerts in order to notify emergency corps members through the mobile application.

\section{Proposed System}

The system developed is in charge of performing a real time monitoring of dynamic risk factors such as meteorological variables, polluting gases and oxygen level measures through a wireless sensor network based on a particular IoT device prototype together with a Web service and a mobile application.

In this sense, the Web service manages all environmental measurements registered by the IoT devices (through Firebase Realtime Database) and the controller of forest fire risks based on fuzzy logic. This Web service is responsible for updating data (environmental measurements, IoT devices, forest fire alerts, etc.) into different interactive elements disposed on its interface. In this way, the interpretation of environmental conditions is made more efficient through linear and bar graphs, heat maps, gauges, environmental alert information panels and interactive maps. The alerts are sent through notifications to the mobile application which also implements a real time operational module for coordinating emergency tasks between emergency corps headquarters and distributed members around the affected forest areas.

\subsection{IoT devices}

\begin{figure}[hb!]
  \centering
    \includegraphics[scale=0.25]{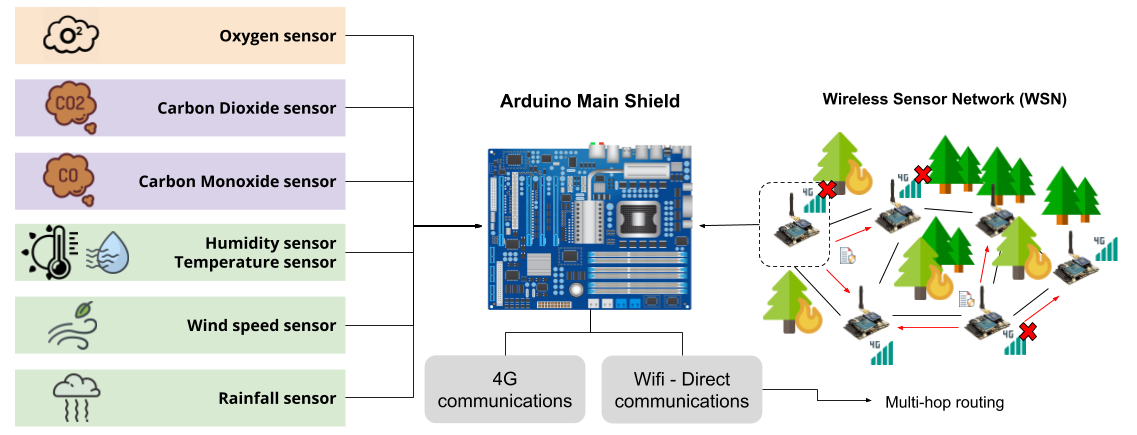}
  \caption{Wireless sensor network architecture}
  \label{fig:arduino}
\end{figure}

Each node in the WSN is able to measure important environmental variables for forest fire prevention, detection and fighting. Arduino platform has been used together with six particular sensors for measuring temperature and humidity, wind speed, rainfall, carbon dioxide, carbon monoxide and oxygen, and a 4G and a Wifi modules have also been assembled. Hence,  communication between every sensor node and the Web service is possible. Apart from that, communication among sensor nodes is also available. GPS service has also been implemented to achieve IoT devices locations (see Figure \ref{fig:arduino}).

Once IoT device is deployed, environmental measurement cycle starts. Firstly, a value of each monitored environmental variable is measured through integrated sensors. New environmental data is processed together with specific IoT device parameters such as location (latitude and longitude), battery level and the International Mobile Station Equipment Identity (IMEI). This last private parameter is used to uniquely identify the corresponding device. After data is formatted, it is signed through Lamport's signature scheme and encrypted by using AES encryption algorithm together with Cipher Block Chaining (CBC) mode. After new measured environmental information is sent,  the Web service allows to tune up the frequency to receive information from each IoT devices. The default time among every measurement cycle has been considered by default as 5 minutes but it can be tunned depending on the severity of forest fire risks.

Wifi module is assembled to the main Arduino shield allowing Wifi-Direct \cite{funai2015supporting} communications among IoT nodes. Wifi-Direct makes bidirectional connections between devices easier and allows data transmission through ranges up to 200 meters without needing any access point. Depending on some static factors such as the frequency of wildfire incidents occurrence or the orography, the proposed wireless sensor network has to be designed. Other factor to take into account is human activities nearby forest areas that may affect the ecosystem state.

\subsubsection{Multi-hop routing}

Since it is possible that some forest areas may be located out of network coverage, multi-hop routing has been implemented in order to ensure that new environmental information may always reach the Web service through other nodes used as relays when the specific location of the forest area may be temporarily out of the network coverage.

Peer to peer communication is performed through Wifi-direct. When an IoT device is not able to send the recent measured environmental information through 4G module, it is necessary to establish secure communications with other close available sensor nodes in order to transmit this information to the Web service. Every time a sensor node is not able to communicate with the Web service as result of being out of network coverage, multi-hop routing is executed taking into account these steps:
\begin{itemize}
\item IoT node out of network coverage connects to a close neighbour node and transmits the recent registered environmental information through Wifi-direct.

\item Destination node receives environmental information and checks if network coverage is currently available in its area. Every IoT node keeps a list of environmental information packages previously received. All data packages to be sent to the Web service are signed by its node owner through Lamport's signature scheme.

\item Whether a data package reaches more than one time the same neighbour node, retransmission process stops to avoid congestion in the WSN. 

\item On the other hand, if a data package is received for the first time, it is sent to the Web service, if network coverage is currently available, or retransmitted to other nodes through Wifi-direct, otherwise.
\end{itemize}

\section{System security}

This proposal includes different security mechanisms to provide with secure communications among IoT devices, the Web service and the mobile application. In this sense, relevant security requirements for IoT deployment such as data privacy, confidentiality and integrity together with authentication  have been considered in detail and implemented \cite{khan2017iot}.

\begin{figure}[ht!]
  \centering
    \includegraphics[scale=0.18]{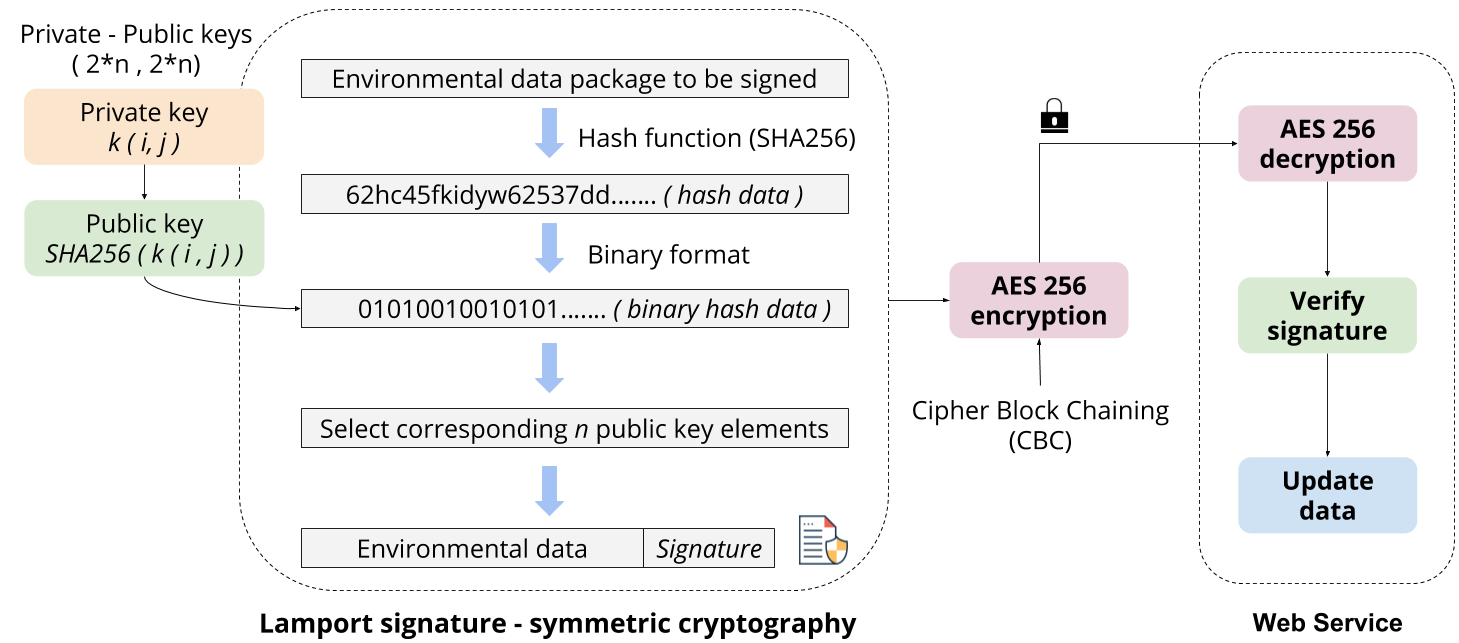}
  \caption{Lamport's signature and AES 256 CBC mode encryption}
  \label{fig:systemsecurity}
\end{figure}

One of the challenges to be considered is the authentication of environmental measurements registered by IoT devices in order to avoid node spoofing and ensuring their integrity. For this purpose, Lamport's signature has been used. This type of signature is based on one-way functions \cite{naor1989universal} and a cryptographic hash function (SHA256). A key pair composed by a private key and a public one are generated. Both contains 2*n elements, but the public one is obtained by applying a hash function on each private key element. After the key pair generation, a hash function is also used on to the content data (measured environmental information together with other device parameters) in order to obtain a specific bits sequence that will be used to select the exact number of bits from the public key and perform Lamport's signature (see Figure \ref{fig:systemsecurity}). 

Hash or Merkle trees are used together with this signature for preventing its one-time use by each key pair generated when one single message is signed, which means many keys should have to be published whether many messages are signed. 

A block cipher algorithm has been implemented for encrypting data packages that will be sent to the Web service from the IoT devices. AES encryption algorithm based on a key of 256 bits together with Cipher Block Chaining (CBC) mode (considered as the current standard encryption) and Zero padding have been implemented without involving an additional cost of time in the environmental measurements management. A key predistribution process has been considered to provide particular keys necessaries to perform this encryption process. 

Robust authentication has been proposed taking into account the  Open Web Application Security Project (OWASP) guidelines. Authentication tokens and secure data transmission protocol through HTTPS are implemented. Regarding passwords management, hash functions have been used and secret keys (such as API keys or encryption algorithm private keys) are also encrypted. A private administration interface with privileged access has been defined in order to manage system resources or relevant metadata registered.

\section{Conclusions}
The main objective of this proposal is to contribute to solve some environmental challenges related to forest fires prevention, detection and tracking through the deployment of wireless sensor networks. It makes possible to monitor dynamic forest fire risk factors in real time. In this sense, a measurement interface through novel IoT technologies, devices and sensors together with multi-hop routing, a forest fire risks controller based on fuzzy logic, a Web service and a mobile application have been implemented. All these elements allow to analyze wildfire incidents and to generate forest fire risk alerts providing an enhancement in response time of emergency corps. On the other hand, security mechanisms are proposed ensuring integrity and confidentiality of information through an encrypting process based on AES 256 and Lamport's signature.

Multiple research lines remain open, such as improving the fuzzy-based controller of forest fire environmental risks through introducing machine learning and new interesting environmental sensors. Regarding the use of machine learning, historical weather data may be used for initial generation of training sets since they include the input variables included in the model presented here.The possibility of implementing a secure wireless sensor network including  Blockchain technology deserves to be studied. Finally, improving wireless sensor network distribution through interesting forest areas according to static risk factors such as the orography or the widlfire incidents frequency should also be considered.

\section{Acknowledgements}
Research supported by Binter-Sistemas grant and the Spanish Ministry
of Economy and Competitiveness, the European FEDER Fund, and the CajaCanarias Foundation, under Projects TEC2014-54110-R, RTC-2014-1648-8, MTM2015- 69138-REDT,TESIS- 2015010106 and DIG02-INSITU.

\bibliographystyle{IEEEtran}
\bibliography{bibliography}
\end{document}